\documentstyle[pra,aps,preprint]{revtex}
\draft
\begin{document}
\author{Dmitri Kilin,
Ulrich Kleinekath\"ofer, and  
Michael Schreiber}
\address{
  Institut f\"ur Physik, 
  Technische Universit\"at, D-09107 Chemnitz, Germany }
\title{Electron Transfer in Porphyrin  Complexes 
in Different Solvents}
\date{\today}
\maketitle
\begin{abstract} 
  The electron transfer in different solvents is investigated for systems
  consisting of donor, bridge and acceptor.  It is assumed that vibrational
  relaxation is much faster than the electron transfer.  Electron transfer
  rates and final populations of the acceptor state are calculated
  numerically and in an approximate fashion analytically.  In wide
  parameter regimes these solutions are in very good agreement.  The theory
  is applied to the electron transfer in ${\rm H_2P-ZnP-Q}$ with free-base
  porphyrin (${\rm H_2P}$) being the donor, zinc porphyrin (${\rm ZnP}$)
  the bridge, and quinone (${\rm Q}$) the acceptor.  It is shown that the
  electron transfer rates can be controlled efficiently by changing the
  energy of the bridging level which can be done by changing the solvent.
  The effect of the solvent is determined for different models.
\end{abstract}


\section{Introduction \protect \label{chem-intr}}
Electron transfer (ET) is a very important process in biology, chemistry
and physics \cite{jort99,d1,b8,n11,b1}.  The most well known ET theory is
the one of Marcus \cite{marc56}.  Of special interest is the ET in
configurations where a bridge (B) between donor (D) and acceptor (A)
mediates the transfer.  On this kind of ET we will focus in this paper.
The primary step of ET in bacterial photosynthetic reaction centers is of
this type \cite{bixo91} and a lot of work in this direction was done after
the structure of the protein-pigment complex of the photosynthetic reaction
centers of purple bacteria was clarified in 1984 \cite{deis84}. Many
artificial systems especially self-organized porphyrin complexes have been
developed to model this bacterial photosynthetic reaction center
\cite{b8,w1,j1}.

Bridge-mediated ET reactions can occur via different mechanisms
\cite{n11,d2,s1,s7}: incoherent sequential transfer in which the bridge
level is populated or coherent superexchange \cite{m6,k8} in which the
mediating bridge level is not populated but nevertheless necessary for the
transfer.  Changing a building block of the complex \cite{w1,r4,z4} or
changing the environment \cite{r4,b7} can modify which mechanism is mainly
at work.  Actually, there is still a discussion in literature whether
sequential transfer and superexchange are limiting cases of one process
\cite{sumi96} or whether they are two processes which can coexist
\cite{bixo91}.  To clarify which mechanism is present in an artificial
system one can systematically vary the energetics of the complex.  In
experiments this is done by substituting parts of the complexes
\cite{w1,j1,r4,z4,p2} or by changing the polarity of the solvent \cite{r4}.
Also the geometry and size of the bridge block can be varied, and in this
way the length of the subsystem through which the electron has to be
transfered \cite{w1,d2,p2,h1,l2,m10} can be changed.

Superexchange occurs due to coherent mixing of the three or more states of
the system \cite{m6,k8,r2,e1}.  The ET rate in this channel depends
algebraically on the differences between the energy levels \cite{w1,j1} and
decreases exponentially with increasing bridge length \cite{m6,m10,e1}.
When incoherent effects such as dephasing dominate the transfer is mainly
sequential \cite{s1,m10}, i.~e., the levels are occupied mainly in
sequential order \cite{b1,s1,s7,r4}.  The dependence on the differences
between the energy levels is exponential \cite{w1,j1}.  An increase of the
bridge length induces only a small reduction in the ET rate
\cite{b1,h1,m10,e1,p3}. This is why sequential transfer is the desired
process in molecular wires \cite{m10,davi98}.

In the superexchange case the dynamics is mainly Hamiltonian and can be
described on the basis of the Schr\"odinger equation. The physically
important results can be obtained by perturbation theory \cite{m6,c3} and,
most successfully, by the semiclassical Marcus theory \cite{marc56}.  The
complete system dynamics can be directly extracted by numerical
diagonalization of the Hamiltonian \cite{m10,j22}.  In case of sequential
transfer the influence of an environment has to be taken into account.
There are quite a few different ways how to include an environment modeled
by a heat bath.  The simplest phenomenological descriptions are based on
the Einstein coefficients or on the imaginary terms in the Hamiltonian
\cite{weis99,l4}, as well as on the Fokker-Planck or Langevin equations
\cite{weis99,l4}.  The most accurate but also numerically most expensive
way is the path integral method \cite{weis99}. This has been applied to
bridge-mediated ET especially in the case of bacterial photosynthesis
\cite{pim}.  Bridge-mediated ET has also been investigated using Redfield
theory \cite{s7,felt95}, by propagating a density matrix (DM) in Liouville
space \cite{s1} and other methods (e.~g.\ 
\cite{e1,j22,lin95,kueh96,guo94}).

The purpose of the present investigation is to present a simple,
analytically solvable model based on the DM formalism \cite{f3,blum96} and
apply it to a porphyrin-quinone complex which is taken as a model system
for the bacterial photosynthetic reaction center.  The master equation
which governs the DM evolution as well as the appropriate relaxation
coefficients can be derived from such basic informations as
system-environment coupling strength and spectral density of the
environment \cite{f3,blum96,r1,m11,kuhn94,j3,m11-n1,k5,k5-s1}.  In the present model
relaxation is introduced in a way similar to Redfield theory but in site
representation not in eigenstate representation.  A discussion of
advantages and disadvantages these representations has been given elsewhere
\cite{dav98a}.  
The equations for the DM are the same as in the generalized stochastic
Liouville equation (GSLE) model \cite{cape85,cape94} for exciton transfer 
which is an extension of the
Haken-Strobl-Reineker (HSR) model \cite{rein82,hake72}
to a model with a quantum bath.
Here we give an analytic solution to these equations. 
The present equations for the DM obtained are also similar to those of Ref.\ 
\cite{d2} where relaxation is introduced in a phenomenological fashion but
only a steady-state solution is found in contrast to the model introduced
here.  In addition the present model is applied to a concrete system.  A
comparison of the ET time with the bath correlation time allows us to
regard three time intervals of system dynamics: the interval of memory
effects, the dynamical interval, and the kinetic, long-time interval
\cite{rein82}.  In the framework of DM theory one can describe the ET
dynamics in all three time intervals. However, often it is enough to find
the solution in the kinetic interval for the explanation of experiments
within the time resolution of most experimental setups, as has been done in
Ref.~\cite{d2,w4}.  The master equation is analytically solvable only for
simple models, for example \cite{l4,k9}.  Most investigations are based on
the numerical solution of this equation \cite{s1,p3,m11,m11-n1}.
Here we perform numerical as well as approximate analytical calculations
for a simple model.  Since the solution can be easily obtained, the
influence of all parameters on the ET can be examined.

The paper is organized as follows.  In the next section we introduce the
model of a supermolecule which we use to describe ET processes.  The
properties of an isolated supermolecule are modeled in subsection
\ref{chem-isolate}, as well as the static influence of the environment.
The dynamical influence of bath fluctuations is discussed and modeled by a
heat bath of harmonic oscillators in section \ref{chem-SB}.  The reduced DM
equation of motion (RDMEM) describing the excited state dynamics is
presented in subsection \ref{chem-RDM}.  In subsection \ref{chem-scaling}
the system parameter dependence on the solvent dielectric constant is
discussed for different models of solute-solvent interaction.  In
subsection \ref{chem-parameters} system parameters are determined.  The
methods and results of the numerical and analytical solutions of the RDMEM
are presented in section \ref{chem-results}.  The dependencies of the ET
rate and final acceptor population on the system parameters are given for
the numerical and analytical solutions in subsection \ref{chem-super}.  The
analysis of the physical processes in the system is also performed there.
In subsection \ref{chem-solvents} we discuss the dependence of the ET rate
on the solvent dielectric constant for different models of solute-solvent
interaction and compare the calculated ET rates with the experimentally
measured ones.  The advantages and disadvantages of the presented method in
comparison with the GSLE model \cite{cape85,cape94} and the method of Davis
et~al.~\cite{d2} are analyzed in subsection \ref{chem-Davis}.  In the
conclusions the achievements and possible extensions of this work are
discussed.

\section{Model \protect \label{chem-model}}
\subsection{System Part of the Hamiltonian\label{chem-isolate}}
The photoinduced ET in supermolecules consisting of three sequentially
connected molecular blocks, i.~e., donor, bridge, and acceptor, ($M=1,2,3$)
is analyzed. The donor is not able to transfer its charge directly to
acceptor because of their spatial separation. Donor and acceptor can
exchange their charges only through B.  In the present investigation the
supermolecule consists of free-base tetraphenylporphyrin (${\rm H_2P}$) as
donor, zinc substituted tetraphenylporphyrin (${\rm ZnP}$) as bridge, and
p-benzoquinone as acceptor \cite{r4}. In each of those molecular blocks we
consider only two molecular orbitals ($m=0,1$), the highest occupied
molecular orbital (HOMO) and the lowest unoccupied molecular orbital (LUMO)
\cite{gout61}.  Each of these orbitals can be occupied by an electron or
not, denoted by $|1\rangle$ or $|0\rangle$, respectively.  This model
allows us to describe four states of each molecular block, the neutral
ground state $|1\rangle_{\rm HOMO}|0\rangle_{\rm LUMO}$, the neutral
excited state $|0\rangle_{\rm HOMO}|1\rangle_{\rm LUMO}$, the positively
charged ionic state $|0\rangle_{{\rm HOMO} }|0\rangle_{\rm LUMO}$, and the
negatively charged ionic state $|1\rangle_{\rm HOMO}|1\rangle_{\rm LUMO}$.
$c^+ _{M m} = | 1 \rangle_{M m} \langle 0 | _{M m}$, $c _{M m} = | 0
\rangle_{M m} \langle 1 | _{M m}$, and $\hat n _{M m} = c^+_{M m} c _{M m}$
describe the creation, annihilation, and number of electrons in orbital ${M
  m}$, respectively, while $\hat n _M = \sum_m \hat n_{M m}$ gives the
number of electrons in a molecular block.  The number of particles in the
whole supermolecule is conserved, $\sum_M {\hat n}_M = const$.

Each of the electronic states has its own vibrational substructure.  As a
rule for the porphyrin-containing systems the time of vibrational
relaxation is two orders of magnitude faster than the characteristic ET
time \cite{r4}.  Because of this we assume that only the ground vibrational
states play a role and we do not include the vibrational substructure.  A
comparison of the models with and without vibrational substructure has been
given elsewhere \cite{schr99}.

Below we consider the evolution of single charge-transfer exciton states in
the system.  For the full description of the system one also should include
photon modes to describe for example the fluorescence from the LUMO to the
HOMO in each molecular block transferring an excitation to the
electro-magnetic field.  But the rates of fluorescence and recombination
are small in comparison to other processes for porphyrin-type systems
\cite{r4,fras94}.  When fluorescence does not have to be taken into
account, all states except $\left| {\rm D^*B A } \right>$ ($ M = 1$),
$\left| {\rm D^+B^-A } \right>$ ($ M = 2$), and $\left| {\rm D^+B A^-}
\right>$ ($ M = 3$) remain essentially unoccupied, while those three take
part in the intermolecular transport process.  In this case the number of
states coincides with the number of sites in the system and we label the
states $\mu{}=1, 2, 3$ instead of $\{ M, m \}$.

For the description of the ET and other dynamical processes in the system
placed in a dissipative environment we introduce the Hamiltonian
\begin{eqnarray}
\label{1}
\hat H = \hat H_{\rm S} + \hat H_{\rm B} + \hat H_{\rm SB}~,
\end{eqnarray}
where $\hat H_{\rm S}$ describes the supermolecule, $\hat H_{\rm B}$ the
dissipative bath, and $\hat H_{\rm SB}$ their interaction.  We are
interested in the kinetic limit of the excited state dynamics here.  For
this limit we assume that the relaxation of the solvent takes only a very
short time compared to the system times of interest.

The effect of the solvent is twofold.  On one hand the system dynamics is
perturbed by the solvent state fluctuations, independent of the system
states.  $\hat H_{\rm SB}$ shall only reflect the dynamical influence of
the fluctuations leading to dissipative processes as discussed in the next
subsection.  On the other hand the system states are shifted in energy
\cite{a15},
\begin{equation}
  \label{2-Hami}
  \hat H_{\rm S} = \hat H_0 + \hat H_{\rm es} + \hat V,
\end{equation}
due to the static 
influence of the solvent 
which is determined by the relaxed value 
of the solvent polarization and in general 
also includes the non-electrostatic contributions such as 
van-der-Waals attraction, 
short-range repulsion, 
and hydrogen bonding
\cite{Georgievski,chri99}.
In Eq.~(\ref{2-Hami}) the energy of free and noninteracting blocks 
$\hat
    H_0
         =
             \sum_{M m}
                 E_{M m} 
                 \hat 
                    n_{M m}$, 
is given by 
the energies $E_{M m}$ of orbitals 
$M m$     in the independent electron
          approximation \cite{n11,kili98b}.  
The $E_{M m}$ are chosen 
to reproduce 
the ground-state--excited-state transitions 
e.~g.~${\rm D} \rightarrow {\rm D}^*$, which
change only a little for different solvents \cite{r4}
and are assumed to be  constants here.
To determine $E_{M m}$ 
one starts from fully ionized
double bonds 
in each molecular block \cite{kili98b}, calculates the one-particle
states 
and fills these
orbitals 
with
two electrons each 
starting from the lowest
energy.
By exciting, removing, adding the last 
electron to the model system
one  obtains
the energy of the excited, oxidized, reduced
molecular block 
in the independent particle
 approximation.

The inter-block hopping term 
$$\hat V = \sum_{\mu{}\nu} v_{\mu{}\nu} (\hat V^+_{\mu{}\nu} + \hat V_{\mu{}\nu}) \left[
  (\hat n_\mu{}-1)^2+(\hat n_\nu -1)^2 \right]$$ in Eq.~(\ref{2-Hami}) includes
the hopping operators $ \hat V_{\mu{}\nu} = c^+_{N 1} c_{M 1}, $
and 
the coherent couplings  $v_{\mu{}\nu}$.
We assume $v_{13}=0$ because there is no
direct connection between donor and acceptor.
The scaling of $v_{\mu{}\nu}$ for different solvents
is discussed in subsection \ref{chem-scaling}.

The electrostatic interaction $\hat H_{\rm es}$ 
scales  
like energies of a system of charges in a single or multiple 
cavity surrounded by 
a medium  with static dielectric constant $\epsilon_{\rm s}$
according to the classical reaction field theory \cite{Boettcher}.
Here we consider two models of scaling.
In the first model each molecular block
is in an individual cavity in the dielectric.
For this case the electrostatic energy reads
\begin{equation}
  \hat H_{\rm es} = S^H(\epsilon_{\rm s}) \left( \hat H_{\rm el} + \hat
    H_{\rm ion} \right).
  \label{static-SB-inter}
\end{equation}
$$ \hat H_{\rm el} 
                  =   \sum_\mu{}
                          |\hat n_\mu{}-1|
                               {e^2}
                               \left(
                                     {4 \pi \epsilon_0}
                                     {r_\mu{}}
                               \right)^{-1} 
$$ 
takes the electron interaction into account while bringing an
additional charge onto the block $\mu{}$ and thus describes 
the energy to create an isolated ion.  
This term depends on the characteristic radius  $r_\mu{}$ 
of the molecular block. 
The interaction between the  ions
$$  \hat H_{\rm ion}
                   =    \sum_\mu{} 
                        \sum_{\nu} (\hat n_\mu{}-1)
                                   (\hat n_{\nu}-1)
                                        {e^2}
                                        \left(
                                           {4\pi\epsilon_0}
                                           {r_{\mu{}\nu}}
                                        \right)^{-1}
$$  
 depends on the distance between 
the molecular blocks $r_{\mu{}\nu}$.  
Both distances $r_\mu{}$ and $r_{\mu{}\nu}$
 are also used in  Marcus theory \cite{marc56}.  
The term $H_{\rm el}+H_{\rm ion}$ reflects the interaction of charges 
inside the supermolecule 
which is weakend by the reaction field  according 
to the Born formula \cite{Karelson} 
\begin{equation}
\label{Born-scaling}
S^H
       =    
              1   +   \frac
                           {1-\epsilon_{\rm s}}
                           {2\epsilon_{\rm s}}. 
\end{equation}

In the second model, considering the supermolecule as one object 
placed in a single cavity  of  constant radius 
one has to use the Onsager term \cite{Karelson}.
This term
is state selective,  it gives a contribution only for the states
with nonzero dipole moment, i.e., charge separation.  Defining the
static dipole moment operator as
$\hat{\vec{p}}=\sum\limits_{\mu{}\nu}(\hat n_\mu{}-1)(\hat
n_\nu-1)\vec r_{\mu{}\nu}e$ 
we obtain 
$
\hat H
_{\rm es}
           =   
                S^H 
                    \hat{ \vec{p} }^2
                /   {r_{13}}$,
with Onsager scaling
\begin{eqnarray}
S^H      
          &=&
                \frac{1-\epsilon_{\rm s}}
                     {2\epsilon_{\rm s}+1}. 
\label{Onsager-scaling}
\end{eqnarray}

\subsection{Microscopic Motivation of the System-Bath Interaction and the Thermal Bath\label{chem-SB}}
One can express
the dynamic part of the system-bath interaction as
\begin{eqnarray}
\hat H
_{\rm SB}
           =    
                 -    \int d^3 \vec{r} 
                      \sum_{\mu{}\nu} 
                          \hat{\vec{D}}_{\mu{}\nu}(\vec r)
                          \cdot{}
                          \Delta 
                          \hat{\vec {P}}(\vec r).
\label{dynam-SB-inter}
\end{eqnarray}
Here $\hat{\vec {D}}_{\mu{}\nu}(\vec r)$ denotes the field of the
electrostatic displacement at point $\vec r$ induced by the system
transition dipole moment 
$\hat{\vec{p}}_{\mu{}\nu}
            =  \vec{p}_{\mu{}\nu}
               (\hat{V}^+_{\mu{}\nu}+\hat V_{\mu{}\nu} )$ \cite{l4}.
The field of the environmental polarization is denoted as $\hat
{\vec{P}}(\vec{r}) = \sum_n \delta(\vec{r}-\vec{r}_n)
\hat{\vec{d}}_n$, where $ \hat{\vec{d}}_n$ is the $n$th 
dipole of the environment and $\vec r_n$ its position.
Only fluctuations of the environment polarization
$\Delta \hat{\vec{P}}(\vec r)$ influence the system dynamics.
Averaged over the angular dependence the interaction reads \cite{a15}
\begin{eqnarray}
\label{averaged_value_of_interaction}
\hat H
_{\rm SB}      =
                  -  \sum 
                     \limits_{\mu{}\nu n}
                     \frac{1}{4 \pi \epsilon_0}
                     \left(
                         \frac{2}{3}
                     \right)^{\frac{1}{2}}
                     \frac
                          {|\hat {\vec p}_{\mu{}\nu}|\Delta|\hat{\vec d_n}|}
                          {|\vec r_n|^3}.
\end{eqnarray}

The dynamical influence of the solvent is described with a thermal bath
model.  The deviation $\Delta \left| \hat{\vec{d}}_n\right|$ of $d_n$ from
its mean value is determined by temperature induced fluctuations.  For
unpolar solvents described by a set of harmonic oscillators the
diagonalization of their interaction yields a bath of harmonic oscillators
with different frequencies $\omega_\lambda$ and effective masses
$m_\lambda$.  In the case of a polar solvent the dipoles are interacting
rotators as, e.g.  used to describe magnetic phenomena
\cite{yosi96,tyab67}.  The elementary excitation of each frequency can
again be characterized by an appropriate harmonic oscillator.  So we use
generalized coordinates of solvent harmonic oscillator modes $ \hat
Q_\lambda = \sqrt{ {\hbar} \left( {2m_\lambda\omega_\lambda} \right)^{-1} }
(\hat a_\lambda + \hat a^+_\lambda) $ for polar as well as unpolar
solvents.  The occupation of the $i$th state of the $\lambda$th oscillator
is defined by the equilibrium DM $ \rho _{\lambda, ij } = \exp{ \left[ -
    {\hbar \omega _\lambda i} /(k_{\rm B}T) \right]} \delta_{ij}$.

All mutual orientations and distances of solvent molecules have equal
probability.  An average over all spatial configurations is performed.  The
interaction Hamiltonian~(\ref{averaged_value_of_interaction}) is written in
a form which is bilinear in system and bath operators:
\begin{eqnarray}
  \hat H_{\rm SB} = \left[ \sum_{\mu{}\nu} p_{\mu{}\nu} (\hat V_{\mu{}\nu} +\hat
    V^+_{\mu{}\nu} ) \right] \left[ \sum_\lambda K_\lambda (\hat a^+_\lambda +
    \hat a_\lambda) \right] S_{\rm SB}
\label{phase}
\end{eqnarray}
$p_{\mu{}\nu} K_{\lambda} $ denotes the interaction intensity between the bath
mode $a_{\lambda}$ of frequency $\omega_{\lambda}$ and the quantum
transition between the LUMOs of molecules $\mu{}$ and $\nu$ with frequency
$\omega_{\mu{} \nu } = \left( E_{\mu{}}-E_{\nu } \right) / \hbar$.  The scaling
function $S_{\rm SB}$ reflects the properties of the solvent.  Explicit
expressions for the solvent influence are still under discussion in the
literature \cite{Georgievski,chri99}.

\subsection{Reduced Density Matrix Approach\label{chem-RDM}}
The interaction of the system with the bath of harmonic oscillators
describes the irradiative energy transfer from the system to the solvent as
modeled by Eq.~(\ref{phase}).  For the description of the dynamics we use
the reduced DM which can be obtained from the full DM $\rho$ by tracing
over the environmental degrees of freedom $\sigma={\rm Tr_B}\rho$
\cite{blum96} with the evolution operator technique \cite{b7,a16},
restricting ourselves to the second order cumulant expansion \cite{y5}.
Here we apply the Markov approximation, i.e.,~we restrict ourselves to the
limit of long times.  Furthermore, we replace the discrete set of bath
modes with a continuous one.  To do so one has to introduce the spectral
density of bath modes $J ( \omega ) = \pi \sum_\lambda K_\lambda^2 \delta(
\omega - \omega_\lambda)$.
Finally one obtains the following master equation
\begin{eqnarray}
  \dot \sigma _{\kappa \lambda} = && - \frac{i}{\hbar} \left([ \hat H_{\rm
      S}, \sigma ]\right)_{\kappa \lambda} + 2\delta_{\kappa \lambda }
  \sum\limits_\mu{} \left\{ \Gamma_{\mu{}\kappa} \left[ n(\omega_{\mu{}\kappa })+1
    \right] + \Gamma_{\kappa \mu{}} n(\omega_{\kappa \mu{}}) \right\} \sigma_{\mu{}\mu{}}
  \nonumber \\ && - \sum\limits_\mu{} \left\{ \Gamma_{\mu{}\kappa} \left[
      n(\omega_{\mu{}\kappa })+1 \right] + \Gamma_{\kappa \mu{}} n(\omega_{ \kappa
      \mu{}}) + \Gamma_{\mu{}\lambda } \left[ n(\omega_{\mu{}\lambda })+1 \right] +
    \Gamma_{ \lambda \mu{}} n(\omega_{ \lambda \mu{}}) \right\} \sigma_{\kappa
    \lambda } \nonumber \\ && + \left\{ \Gamma_{ \lambda \kappa} \left[
      2n(\omega_{ \lambda \kappa })+1 \right] + \Gamma_{\kappa \lambda }
    \left[ 2n(\omega_{\kappa \lambda })+1 \right] \right\} \sigma_{ \lambda
    \kappa},
\label{RWA-short}
\end{eqnarray}
where $n(\omega)=[\exp{(\hbar\omega/k_B T)}-1]^{-1}$ 
denotes the Bose-Einstein distribution.
The damping  constant
\begin{equation}
\label{g=kv}
\Gamma_{\mu{}\nu} = S^2_{\rm SB} \hbar^{-2} J ( \omega _{ \mu{} \nu } ) p^2_{\mu{}\nu}
\end{equation}
reflects the coupling of the transition $\left| \mu{}\right>\to\left| \nu
\right>$ to a bath mode of the same frequency.  It depends on the density
of bath modes $J$ at the transition frequency $\omega_{\mu{}\nu}$ and on the
transition dipole moments $p_{\mu{}\nu}$.  A RDMEM of similar structure was
used for the description of exciton transfer in the Haken, Strobl, and
Reineker (HSR) model \cite{rein82,hake72} and the generalized stochastic
Liouville equation (GSLE) model \cite{cape85,cape94}.  The HSR method
originating from the stochastic bath model, is valid only in the high
temperature limit \cite{cape94}.  The GSLE method \cite{cape85,cape94}
appeals to the quantum bath model with system-bath coupling of the form
$\hat H_{\rm SB} \sim \hat V^+ \hat V \left( \hat a^+_\lambda + \hat
  a_\lambda \right)$, which modulates the system transition frequency.  In
Ref.~\cite{cape85,cape94} the equations for exciton motion are derived
using the projection operator technique.  Taking the different system-bath
coupling we have derived the RDMEM which coincides with GSLE
\cite{cape85,cape94}.  Both GSLE and our RDMEM are able to describe
correctly finite temperatures.  Below we neglect the last term of
Eq.~(\ref{RWA-short}) corresponding to the $\bar \gamma$ term in the HSR
and GSLE models because the rotating wave approximation (RWA) is applied.

For the sake of convenience of analytical and numerical calculations we
replace $\Gamma_{\mu{}\nu}$ and the population of the corresponding bath mode
$n(\omega_{\mu{} \nu})$ with the dissipative transitions
$d_{\mu{}\nu}=\Gamma_{\mu{}\nu}|n(\omega_{\mu{}\nu})|$ and the corresponding dephasings
$ \gamma _{\mu{}\nu } = \sum_\kappa \left( d_{\mu{}\kappa } + d_{ \kappa \nu }
\right)/2.  $ With this one can express the RDMEM (\ref{RWA-short}) in the
form
\begin{eqnarray} 
\label{tosolve1}
\dot \sigma_{\mu{}\mu{}} &=& -i/\hbar \sum_ \nu (v_{\mu{}\nu } \sigma_{ \nu \mu{}} -
\sigma_{\mu{}\nu } v_{ \nu \mu{}}) - \sum_ \nu d_{\mu{}\nu } \sigma_{\mu{}\mu{}} + \sum_ \nu
d_{ \nu \mu{}} \sigma_{ \nu \nu }, \\ 
\label{tosolve2}
\dot \sigma_{\mu{}\nu } &=& \left( - i\omega_{\mu{}\nu } - \gamma_{\mu{}\nu } \right)
\sigma_{\mu{}\nu } - i/\hbar v_{\mu{}\nu } (\sigma_{ \nu \nu }-\sigma_{\mu{}\mu{}}).
\end{eqnarray} 
The parameters controlling the transitions between the selected states are
discussed in subsection~\ref{chem-parameters}.

\subsection{Scaling of the Damping Constants\label{chem-scaling}}
The relaxation coefficients Eq.~(\ref{g=kv}) include the second power of
the scaling function $S_{\rm SB}$ because one constructs the relaxation
term of Eq.~(\ref{RWA-short}) with the second power of the interaction
Hamiltonian.  The physical meaning of $H_{\rm SB}$ is similar to the
interaction of the system dipole with a surrounding media.  That is why it
is reasonable to use the Onsager expression~(\ref{Onsager-scaling}) for
$S_{\rm SB}$.  In the work of Mataga, Kaifu, and Koizumi~\cite{Mataga} the
interaction energy between the system dipole and the media scales in
leading order as
\begin{equation}
\label{Mataga-scaling}
S_{\rm SB} = - \left[ \frac{2(\epsilon_{\rm s}-1)} {2\epsilon_{\rm s}+1} -
  \frac{2(\epsilon_\infty-1)} {2\epsilon_\infty+1} \right],
\end{equation}
where $\epsilon_\infty$ denotes the optical dielectric constant.  From a
recent paper of Georgievskii, Hsu, and Marcus \cite{Georgievski} we extract
$\Gamma \sim \frac{1}{\epsilon_{\rm s}} -\frac{1}{\epsilon_\infty}$ for the
multiple cavities model assuming $\epsilon_\omega=\epsilon_\infty$.  In
terms of a scaling function it can be expressed as
\begin{eqnarray}
\label{Marcus-scaling}
S_{\rm SB} = \left( {1}/{\epsilon_{\rm s}} - {1}/{\epsilon_\infty}
\right)^\frac{1}{2}.
\end{eqnarray}

As we have already argued in \cite{schr99} the coherent coupling $v_{\mu{}\nu}$
between two electronic states scales with $\epsilon_{\rm s}$ and
$\epsilon_{\infty}$ too, because a coherent transition in the system is
accompanied by a transition of the environment state which is larger for
solvents with larger polarity. As discussed above we neglect the
vibrational substructure of each electronic state because the vibrational
relaxation is about two orders of magnitude faster than the characteristic
ET time. But in contrast to the model with vibrational substructure the
present model does not involve any reaction coordinate.  To reproduce the
results of the more elaborate model with vibrational substructure one has
to scale the electronic couplings $v_{\mu{}\nu}$ with the Franck-Condon overlap
elements $ F_{\rm FC}(\mu{}, 0, \nu, 0) $ between the vibrational ground states
of each pair of electronic surfaces
\begin{equation}
  v_{\mu{}\nu} = v^0_{\mu{}\nu} F_{\rm FC}(\mu{}, 0, \nu, 0)~,
\label{v_scale}
\end{equation}
where $v^0_{\mu{}\nu}$ is the coupling of electronic states of the isolated
molecule. For the calculation of the Franck-Condon factors one has to
introduce the leading (mean) environment oscillator frequency $\omega_{\rm
  vib}$.  Here $\omega_{\rm vib}=1500~{\rm cm}^{-1}$ is used which is
similar to the frequency of the C-C stretching mode.
With this scaling one implicitly introduces a reaction coordinate into
the model.

\subsection{Model Parameters}\label{chem-parameters}
The dynamics of the system is controlled by the following
parameters: energies of system states  $E_\mu{}$, coherent couplings
$v_{\mu{}\nu}$, and  damping  constants  $\Gamma_{\mu{}\nu}$. 

On the basis of the spectral data \cite{r16} and taking reference energy
$E_{\rm DBA}=0$ we determine $E_{\rm D^*BA}=1.82$~{\rm eV} (in ${\rm
  CH_2Cl_2}$).  We take the energy of the state with ET to ${\rm Q}$ from
reference \onlinecite{r4}: $E_{\rm D^+BA^-}=1.42~ {\rm eV}$
\cite{footnote}.  Further Rempel et al.~\cite{r4} estimate the coupling of
initially excited and charged bridge states $\langle {\rm D^*BA} |H| {\rm
  D^+B^-A} \rangle = v^0_{12} = 65~{\rm meV} = 9.8\times{}10^{13}~ ~{\rm s}^{-1}$
and the coupling of the two states with charge separation $\langle {\rm
  D^+B^-A} |H| {\rm D^+BA^-} \rangle = v^0_{23} = 2.2~{\rm meV} =
3.3\times{}10^{12}~ ~{\rm s}^{-1}$.  The values of the couplings are essentially
lower than the energy differences between the relevant system states
\begin{equation}
\label{inequality_w>v}
\hbar \omega_{ij} \gg{}v^0_{ij}.
\end{equation}
This is the reason to remain in site representation instead of eigenstate
representation \cite{davi98}.  The damping constants are found with help of
the analytical solution derived at the end of the next section to be
$\Gamma _{21} = \Gamma _{23} = 2.25 \times{} 10^{12} ~{\rm s}^{-1}$.  The typical
radius of the porphyrin ring is about $r_\mu{}= 5 \pm{}1~{\mathrm \AA{}}$ \cite{z4},
while the distance $r_{\mu{}\nu}$ between the blocks of ${\rm H_2P-ZnP-Q}$
reaches $r_{12}=12.5 \pm{}1~{\mathrm \AA{}}$ \cite{r4,z4}, $r_{23}= 7 \pm{}1~{\mathrm
  \AA{}}$, $r_{13}=14 \pm{}1~{\mathrm \AA{}}$.  The main parameter which controls ET in
a triad is the energy of the state $E_{\rm D^+B^-A}$. This state has a big
dipole moment because of its charge separation and is therefore strongly
influenced by the solvent.  Because of the special importance of this value
we calculate it for the different solvents as a matrix element of the
system Hamiltonian (2).  The calculated values of the energies of the ${\rm
  D^+B^-A}$ state for some solutions are shown in Table~\ref{tab:2}.

\section{Results} \label{chem-results}
The time evolution of the ET in the supermolecule is described by solving
numerically and analytically Eqs.~(\ref{tosolve1})-(\ref{tosolve2}) with
the initial condition of donor excitation with a $\pi$ pulse of appropriate
frequency, i.e., the donor population is set to one.
 
For the numerical simulation we express the system of
Eqs.~(\ref{tosolve1})-(\ref{tosolve2}) in the form
$\dot{\bar\sigma}=A\bar\sigma,$ where $\bar\sigma$ is a vector of dimension
$3^2$ for the model with $3$ system states and the super-operator $A$ is a
matrix of dimension $3^2 \times{}3^2$.  We find an exponential growth of the
acceptor population
\begin{equation}
\label{single-exponent}
P_3(t) = P_3(\infty) \left[ 1-\exp{(-k_{\rm ET}t)} \right],
\end{equation}
where for the solvent MTHF $k_{\rm ET} \simeq 3.59 \times{}10^{8} ~ ~{\rm s}^{-1}$
and $P_3(\infty) \simeq 0.9994$.  The population $P_2$ which corresponds to
charge localization on the bridge does not exceed $0.005$.  This means that
in this case the superexchange mechanism dominates over the sequential
transfer mechanism.  Besides it ensures the validity of characterizing the
system dynamics with $P_3(\infty)$ and
\begin{eqnarray}
\label{fit}
k_{\rm ET} = P_3(\infty) \left\{ \int_0^\infty \left[ 1-P_3(t) \right] dt
\right\}^{-1}.
\end{eqnarray}

The alternative analytical approach 
is performed in the kinetic 
limit 
\begin{equation}
\label{kinetic-limit}
t \gg{}1/    {\rm min}  (\gamma_{ \mu{}\nu }).  
\end{equation}
In Laplace space the inequality (\ref{kinetic-limit}) reads $ s \ll{} {\rm min}
(\gamma_{ \mu{}\nu })$, where $s$ denotes the Laplace variable.  It is
equivalent to replacing the factor $1/(i\omega_{ \mu{}\nu }+\gamma_{ \mu{}\nu }+s)$
in the Laplace transform of Eqs.~(\ref{tosolve1})-(\ref{tosolve2}) with
$1/(i\omega_{ \mu{}\nu }+\gamma_{ \mu{}\nu })$.  This trick allows to substitute
the expressions~(\ref{tosolve2}) for non-diagonal elements of the DM into
Eq.~(\ref{tosolve1}).  After this elimination we describe the coherent
transitions to which the non-diagonal elements contribute by redefinition
of the diagonal RDMEM~(\ref{tosolve1})
\begin{equation}
\label{relax-matrix-new}
\dot \sigma _{\mu{}\mu{}} = - \sum_ \nu g_{ \mu{}\nu }\sigma_{ \mu{}\mu{}} + \sum_ \nu g_{ \nu
  \mu{}}\sigma_{ \nu \nu }.
\end{equation} 
The transition coefficients $g_{ \mu{}\nu }$ 
contain  dissipative and coherent contributions
\begin{equation} 
\label{dissipation}
g_{ \mu{}\nu } = d_{ \mu{}\nu } + v_{ \mu{}\nu }v_{ \nu \mu{}} \gamma_{ \mu{}\nu } \left[
  \hbar^2 \left( \omega^2_{ \mu{}\nu } +\gamma^2_{ \mu{}\nu } \right) \right]^{-1}.
\end{equation} 
Now it is assumed that the bridge is not populated. This allows us to find
the acceptor population in the form of Eq.~(\ref{single-exponent}), where
\begin{eqnarray} 
\label{rate}
k_{\rm ET} = g_{23} + {g_{23}(g_{12}-g_{32})} ({g_{21}+g_{23}})^{-1}, \\ 
\label{population}
P_3(\infty) = {g_{12}g_{23}} \left[ \left( {g_{21}+g_{23}} \right) k_{\rm
    ET} \right]^{-1}.
\end{eqnarray} 
The value of $\Gamma _{\mu{}\nu} = S_{\rm SB}^2 \hbar^{-2} J(\omega_{\mu{}\nu})
p^2_{\mu{}\nu}$ can be found comparing the experimentally determined ET rate
and Eq.~(\ref{rate}).  To calculate $J(\omega_{\mu{} \nu })$ would require a
microscopic model.  To avoid a microscopic consideration we simply take the
same $\Gamma_{\mu{}\nu}$ for all transitions between excited states.  The value
of ET for ${\rm H_2P-ZnP-Q}$ in MTHF is found by Rempel et al. \cite{r4} to
be $k_{\rm ET}=3.6\pm{}.5 \times{}10^8 ~{\rm s}^{-1}$.  If the bridge state has a
rather high energy one can neglect thermally activated processes.  $v_{23}$
is negligibly small with respect to $v_{12}$.  In this case our
result~(\ref{rate}) reads
\begin{eqnarray}
  k_{\rm ET} = v_{12}^2 \Gamma_{21} \Gamma_{23} \left( \hbar^2
    \omega_{21}^2 + \Gamma_{21}^2 \right)^{-1} \left(
    {\Gamma_{21}+\Gamma_{23}} \right)^{-1}.
\end{eqnarray}
With the relation $\Gamma_{21}=\Gamma_{23}$ and the experimental $k_{\rm
  ET}$ one obtains $\Gamma_{21}=\Gamma_{23}\simeq2.25\times{}10^{12} ~{\rm
  s}^{-1}$.  The fit of the numerical solution of
Eqs.~(\ref{tosolve1})-(\ref{tosolve2}) to the experimental $k_{\rm ET}$ in
MTHF gives the same value.  So the damping constants are fixed for a
specific solvent and for other solvents they are calculated with the
scaling functions.  With this method the ET was found to occur with
dominance of the superexchange mechanism with rates $4.6\times{}10^{6} ~{\rm
  s}^{-1}$ for CYCLO and $3.3 \times{}10^8 ~{\rm s}^{-1}$ for ${\rm CH_2Cl_2}$.

\section{Discussion \protect \label{chem-discussion}}
\subsection{Sequential Versus Superexchange\label{chem-super}} 
To discuss how the transfer mechanism depends on the change of parameters
we calculate the system dynamics varying one parameter at a time.  The
dependencies of $k_{\rm ET}$ and
$P_3(\infty)$ on 
$v_{12}$, $v_{23}$ and $\Gamma_{21}$, $\Gamma_{23}$ are shown in Figs. 2
and 3.  The change of each parameter influences the transfer in a different
way.

In particular, $k_{\rm ET}$ depends quadratically on $v_{12}$ from $10^{15}
~{\rm s}^{-1}$ to $10^{12} ~{\rm s}^{-1}$ in Fig.~2.  Below it saturates at
the lower bound $k_{\rm ET} \propto 3 \times{} 10^{5} ~{\rm s}^{-1}$.  This
corresponds to a crossover of the ET mechanism from superexchange to
sequential transfer.  But, due to the big energy difference between donor
and bridge states the sequential transfer efficiency is extremely low.
This is displayed by $P_3(\infty) \simeq 0$.  In the region $v_{12} \approx
v_{13}$ both mechanisms contribute to $k_{\rm ET}$.  The decrease of
$P_3(\infty)$ in this region corresponds to coherent back transfer.  The ET
rate depends on $v_{23}$ in a similar way.  At rather high values of
$v_{12}$, $v_{23} \simeq 10^{15} ~{\rm s}^{-1}$ the
relation~(\ref{inequality_w>v}) is no more valid. For this regime one has
to use eigenstate instead of site representation because the wavefunctions
are no more localized \cite{davi98}.

The variation $\Gamma_{21}$, $\Gamma_{23}$ near the experimental values
shows similar behavior of $k_{\rm ET}(\Gamma_{21})$ and $k_{\rm
  ET}(\Gamma_{23})$ (see Fig. 3).  Here we independently vary $\Gamma_{21}$
and $\Gamma_{23}$.  Both, $k_{\rm ET}(\Gamma_{21})$ and $k_{\rm
  ET}(\Gamma_{23})$ increase linear until the saturation value $7\times{}10^{8}
~{\rm s}^{-1}$ at $\Gamma>10^{12} ~{\rm s}^{-1}$ is reached.  There is
qualitative agreement between the numerical and analytical values.  In
Eq.~(\ref{fit}) infinite time is approximated by $10^{-5} ~{\rm s}$ and so
one cannot obtain ET rates lower than this limit.

The physical meaning of the ET rate dependence on $\Gamma$ seems to be
transparent.  At small values of $\Gamma$ a part of the population
coherently oscillates back and forth between the states. The increase of
the dephasing $\gamma_{\mu{}\nu}$ quenches the coherence and makes the transfer
irreversible.  So transfer becomes faster up to a maximal value.  For the
whole range of $\Gamma$, depopulations $d_{21}$, $d_{23}$ and thermally
activated transitions $d_{12}$, $d_{32}$ always remain smaller than the
coherent couplings, therefore they do not play an essential role.
 
Next, the similarity of the dependencies on $\Gamma_{21}$ and $\Gamma_{23}$
will be discussed on the basis of Eq.~(\ref{rate}).  In the limit $k_B T /
\hbar \omega _{\mu{} \nu} \rightarrow 0$ thermally activated processes with
$\omega_{ \mu{}\nu }<0$ vanish and so $|n(\omega_{ \mu{}\nu })|=0$, while
depopulations with $\omega_{ \mu{}\nu }>0$ remain constant $|n(\omega_{ \mu{}\nu
  })|=1$.  The condition $\omega_{ \mu{}\nu } \gg{}\gamma_{ \mu{}\nu }$ allows us to
neglect $\gamma^2_{ \mu{}\nu }$ in comparison with $\omega^2_{ \mu{}\nu }$.  With
these simplifications
Eq.~(\ref{rate})
becomes
\begin{eqnarray}
\label{symmetric}
k_{\rm ET} \simeq \Gamma_{21}\Gamma_{23} \left( \Gamma_{21} + \Gamma_{23}
\right)^{-1} \left( v_{12}^2/\omega_{21}^2 + v_{23}^2/\omega_{23}^2
\right),
\end{eqnarray}
i.e.~symmetric with respect to 
$\Gamma_{21}$ and $\Gamma_{23}$.

To the largest extent the mechanism of transfer depends 
on the bridge energy  
$E_{\rm D^+B^-A} $
as presented in Fig. 4. 
In different regions one observes 
different types of dynamics. 
For large bridge energy  
$E_{21}=E_{\rm D^+B^-A}-E_{\rm D^*BA} \gg{}0$ 
the numerical and analytical solutions do not differ  
from each other. 
The transfer occurs with the superexchange mechanism. 
The ET rate  
reaches a maximal value of $10^{11}~{\rm s}^{-1}$
for low bridge energies. 
 
While the bridge energy approaches the donor energy the sequential transfer
starts to contribute to the ET process.  The traditional scheme of
sequential transfer is obtained when donor, bridge, and acceptor levels are
arranged in a cascade.  In this region the analytical solution need not
coincide with the numerical solution because the used approximations are no
more valid.  For equal bridge and acceptor energies $k_{\rm ET}$ displays a
small resonance peak in Fig.~4(a).  When the bridge energy is lower than
the acceptor energy the population gets trapped at the bridge.  The finite
$k_{\rm ET}$ for $ E_{21}<E_{31} $ does not mean ET because
$P_3(\infty)\rightarrow~0$.  For the dynamic time interval $t<\gamma_{ \mu{}\nu
  }^{-1}$ a part of the population tunnels force and back to the acceptor
with $k_{\rm ET}$.  The analytical solution~(\ref{rate}) gives a constant
rate for the regime $E_{21}<E_{31}$, while the numerical solution of
Eqs.~(\ref{tosolve1})-(\ref{tosolve2}) is instable, because such coherent
oscillations of population cannot be described by
Eq.~(\ref{single-exponent}) and $k_{\rm ET}$ cannot be fitted with
Eq.~(\ref{fit}).  In Fig.~4 the regime $E_{21}<E_{31}$ occurs for small
$E_{21}$ while $E_{31}$ is kept constant and for large $E_{31}$ while
$E_{21}$ remains constant.

The energy dependence of the final population has a transparent physical
meaning for the whole range of energy.  A large bridge energy ensures the
transition of the whole population to the acceptor.  In the intermediate
case, when the bridge has the same energy as the acceptor, final population
spreads itself over these two states $P_3(\infty)=.5$.  Lowering the bridge
even more the whole population remains on the bridge as the lowest state of
the system.  The dependence of the ET rate on the acceptor energy
$E_{31}=E_{\rm D^+BA^-}-E_{\rm D^*BA}$ in Fig.~4 remains constant while the
acceptor energy lies below the bridge energy.  Increase of $E_{31}$ up to
$E_{21}=1.36~ {\rm eV}$ gives the maximal $k_{\rm ET}\propto\Gamma_{21}$.
When $E_{31}$ increases further the acceptor becomes the highest level in
the system and therefore the population cannot remain on it.
 
\subsection{Different Solvents\label{chem-solvents}}
For the application of the results to various solvents and comparison with
experiment one should use the scaling for energy, coherent couplings, and
damping constants as discussed above.  The combinations of the energy
scaling in subsection \ref{chem-isolate} and damping constants scalings in
subsection~\ref{chem-scaling} are represented in Fig.~5.  An increase in
$\epsilon_{\rm s}$ from $2$ to $4$ leads to an increase of the ET rate, no
matter which scaling is used.  Further increase of $\epsilon_{\rm s}$
induces saturation for the Onsager-Mataga scaling and even a small decrease
Within the applied approximations an increase in the solvent polarizability
and, consequently, of its dielectric constant lowers the bridge and
acceptor energies and increases the system-bath interaction and,
consequently, the relaxation coefficients.  It induces a smooth rise of the
ET rate for the Onsager-Mataga scaling.  On the other hand large
$\epsilon_{\rm s}$ leads to essentially different polarisational states of
the environment for the supermolecule states with different dipole moment.
This reduces the coherent couplings, see Eq.~(\ref{v_scale}), leading for
the Born-Marcus scaling to a small decrease of $k_{\rm ET}$ for large
$\epsilon_{\rm s}$.  The ET rate with this scaling comes closer to the
experimental value $k_{\rm ET}(\epsilon_{\rm s}^{\rm CH_2Cl_2})$.  This
gives a hint that the model of individual cavities for each molecular block
is closer to reality than the model with a single cavity for the whole
supermolecule.

Below we consider Born scaling Eq.\ (\ref{Born-scaling}) for the system
energies and Marcus scaling Eq.\ (\ref{Marcus-scaling}) for the damping
constant to compare the calculated ET rates with the measured ones.  For
the solvents CYCLO, MTHF, and ${\rm CH_2Cl_2}$ one obtains the relative
bridge energies $E_{21}=1.77~ {\rm eV}$, $1.36~ {\rm eV}$, and $1.30~ {\rm
  eV}$, respectively.

The calculated ET rate coincides with the experimental value
\cite{r4} for ${\rm H_2P-ZnP-Q}$ in CYCLO, see table~\ref{tab:2}.
For ${\rm CH_2Cl_2}$ the numerical ET rate 
is approximately 30\%
faster than the experimental value.  
It has to be noted that a value for the damping rates can be chosen such that
the calculated curve almost passes through all three experimental error bars.
On the other hand an error in the present calculation  could be 
due to
(i)    absence of
       vibrational substructure 
       of the electronic states in the present model;
(ii)   incorrect dependence of system states energies on the solvent
       properties; 
(iii)  opening of additional
       transfer channels not mentioned in the scheme shown in Fig.~1.  
Each of these possibilities  needs some comments.

{\it ad} (i): The incorporation of the vibrational substructure will result
in a complication of the model \cite{s7,schr99} giving a more complicated
ET rate dependence on the energy of the electronic states and dielectric
constant.  It should yield the maximal ET rate for nonequal energies of
electronic states, namely for the activationless case when the energy
difference equals the reorganization energy.  For a comparison of the
models with and without vibrational substructure see Ref.~\cite{schr99}.

{\it ad} (ii): Effects as the solvation shell \cite{cich98} do need a
molecular dynamics simulation. The total influence of the solvent is,
probably, reflected in an energy shift between the spectroscopically
observable states $E_{\rm D^*BA}$ and $E_{\rm DB^*A}$ \cite{r4}.

{\it ad} (iii): A solvent with large $\epsilon_{\rm s}$ can bring
high-lying system states closer to the ones included in Fig.~1.
E.g.,~because of its larger dipole moment, $\left| {\rm D^-B^+A} \right>$
is strongly influenced by the solvent.


\subsection{Comparison with similar theories\label{chem-Davis}}
As discussed above the RDMEM are very similar to those of the GSLE model.
This is an extension of the HSR theory in which a  classical bath is
used and for which analytical solutions are available \cite{rein82,cape94}.
We are not aware of any analytical solution of the GSLE model as
presented here. Also this model has not been applied to similar ET processes.

The numerical steady-state method used by Davis et al.\ \cite{d2} is an
attractive one due to its simplicity, but unlike our method it is not able
to give information about the time evolution of the system.  We use a
similar approach derived within a Redfield-like theory.  But we consider
dephasing and depopulation between each pair of levels.  In contrast Davis
et al.\ incorporate relaxation phenomenologically only to selected levels,
dephasing $\gamma$ occurs between excited levels, while depopulation $k$
takes place only for the sink from acceptor to the ground state.  The
advantage of the approach of Davis et al.\ is the possibility to
investigate the ET rate dependence for the bridge consisting of more than
one molecular block.  This was not the goal of the present work but it can
be extended into this direction.  We are interested in the ET in a concrete
molecular complex with realistic parameter values and realistic
possibilities to modify those parameter.  Our results as well as the
results of Davis show that ET can occur as coherent (with the superexchange
mechanism) or dissipative process (with the sequential transfer mechanism).

\section{Conclusions \protect \label{chem-conclusions}}
We have performed a study of the ET in the supermolecular complex ${\rm
  H_2P}-{\rm ZnP}-{\rm Q}$ within the DM formalism.  The determined
analytical and numerical ET rates are in reasonable correspondence with the
experimental data.  The superexchange mechanism of ET dominates over the
sequential transfer.  We have investigated the stability of the model
varying one parameter at a time.  The qualitative character of the transfer
is stable with respect to a local change of system parameter.  The
crossover between the two transfer mechanisms can be induced by lowering
the bridge energy.  The relation of the theory presented here to other
theoretical approaches to ET has been discussed.

The calculations performed in the framework of the present formalism can be
extended in the following directions: (i) Considerations beyond the kinetic
limit.  The vibrational substructure has to be included into the model as
well as solvent dynamics and, probably, non-Markovian RDMEM.  (ii)
Enlargement of the number of molecular blocks in the complex.  (iii)
Initial excitation of states with rather high energy should open additional
transfer channels.

\acknowledgements 
D.~K. thanks U.~Rempel and E.~Zenkevich for stimulating
discussions. Financial support of the DFG is gratefully acknowledged.

\begin{figure}[ht]
\label{chem-schema}
\caption{
Schematic presentation of the energy levels in 
the ${\rm H_2P}-{\rm ZnP}-{\rm Q}$ complex. 
The three states in the boxes play  
the main role 
in ET which  
can happen either sequentially or by a superexchange mechanism. 
Dashed lines refer to sequential transfer,
curved solid line to superexchange, 
dot-dashed to energy transfer followed by ET,
dotted line optical excitation, and
straight solid lines either fluorescence or irradiative recombinations.
}
\end{figure}

\begin{figure}[ht]
\label{chem-vau}
\caption{
Dependence of 
the ET rate            (a) and 
the final acceptor population (b) 
on the coherent couplings
$v_{12}$ (triangles and dashed line, $v_{23}=v_{23}^0=2.2~ {\rm meV}$), 
$v_{23}$ (dots and solid line $v_{12}=v_{12}^0=65~{\rm meV}$).
Symbols represent  numerical solution
of Eqs.~(\ref{tosolve1})-(\ref{tosolve2}),
lines 
analytical solution
(\ref{rate})-(\ref{population}). 
}
\end{figure}

\begin{figure}[ht]
\label{chem-damping}
\caption{
  Dependence of the ET rate on the damping constants $\Gamma_{21}$
  (triangles and dashed line), $\Gamma_{23}$ (dots and solid line).  The
  other parameters correspond to ${\rm H_2P-ZnP-Q}$ in MTHF.  Symbols
  represent numerical and lines analytical solution.  }
\end{figure}

\begin{figure}[ht]
\label{chem-energy}
\caption{
  Dependence of the ET rate (a) and the final acceptor population (b) on
  the energy of B $E_{21}=E_{\rm D^+B^-A}-E_{\rm D^*BA}$ (triangles and
  dashed line, $E_{31}=-0.4~{\rm eV}$) and A $E_{31}=E_{\rm D^+BA^-}-E_{\rm
    D^*BA}$ (dots and solid line, $E_{21}=1.36~{\rm eV}$).  Symbols
  represent numerical and lines analytical solution.  $v_{12}=65~{\rm
    meV}$, $v_{23}=2.2~{\rm meV}$, $\Gamma_{21}=\Gamma_{23}=2.25 \times{}10^{12}~
  ~{\rm s}^{-1}$.  }
\end{figure}

\begin{figure}[ht]
\label{chem-solvent}
\caption{
  The ET rate $k_{\rm ET}$ versus dielectric constant.  The energies of the
  bridge and acceptor scale in accordance with Born
  expression~(\ref{Born-scaling}) (triangles and solid line), Onsager
  expression~(\ref{Onsager-scaling}) (diamonds and dashed line).  Coherent
  couplings and damping constants scale in accordance with Mataga's
  expression~(\ref{Mataga-scaling}) (diamonds and dashed line),
  Georgievskii-Marcus expression~(\ref{Marcus-scaling}) (triangles and
  solid line).  Symbols represent numerical and lines analytical solution.
  Solid crosses with error bars give experimental values \protect
  \cite{r4}. Note that by using a different value of the damping parameter
  $\Gamma$ curves can be calculated which almost pass through all three
  experimental error bars.}
\end{figure}

\begin{table}[] 
  \begin{center} 
    \caption{Energy of the  charged bridge state in 
      different solvents and corresponding ET rates (for calculations Born
      scaling~(\ref{Born-scaling}) of energy and Marcus
      scaling~(\ref{Marcus-scaling}) of dissipation are used). MTHF denotes
      2-methyl-tetrahydrofuran, and CYCLO denotes cyclohexane.}
\begin{tabular}{cccc} 
Solution                                          &  ${\rm CH_2Cl_2}$              & MTHF                          & CYCLO  \\ \hline 
$\epsilon_{\rm s}$,  \cite{r16}                   &  $9.08      $                  & $6.24    $                    & $2.02$ \\ 
$E_{\rm D^+B^-A}$, $ {\rm eV}$                    &  $3.12      $                  & $3.18    $                    & $3.59$ \\   
$  k_{\rm ET},~10^7~{\rm s}^{-1}$, num.           &  $33       $                  & $36      $                    & $0.46$ \\  
$  k_{\rm ET},~10^7~{\rm s}^{-1}$, an.            &  $33       $                  & $36      $                    & $0.46$ \\ 
$  k_{\rm ET},~10^7~{\rm s}^{-1}$, exp. \cite{r4} &  $23 \pm{}5 $                  & $36 \pm{}5$                    & $0+3 $ \\  
     \end{tabular} 
    \label{tab:2} 
  \end{center} 
\end{table}


\begin{thebibliography}{10}
\bibitem{footnote} {One can compare these energies with the energies of the
    same levels of H-Chlorin-Zn-Porphyrin-Quinone in solution of
    butonitrile \cite{j1}, namely $E_{\rm D^*BA}=1.85~ {\rm eV}$, $E_{\rm
      D^+B^-A}=3.05~ {\rm eV}$, $E_{\rm DB^*A}=2.12~ {\rm eV}$, $E_{\rm
      DB+A^-}=2.15~ {\rm eV}$, $E_{\rm D^+BA^-}=1.47~ {\rm eV}$.  The
    present modeling can be applied to this very similar molecular
    aggregate using the same approximations.}
\end{thebibliography}
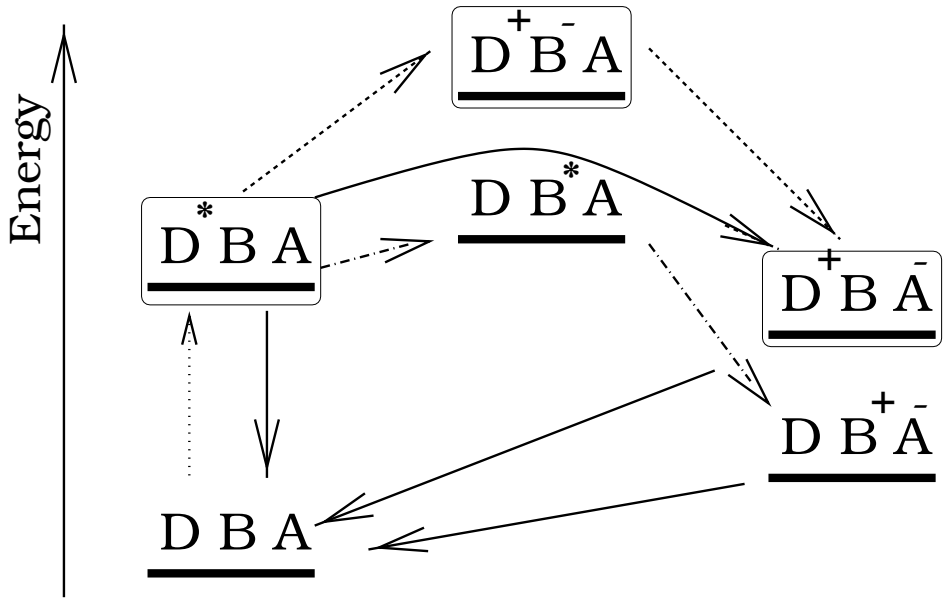
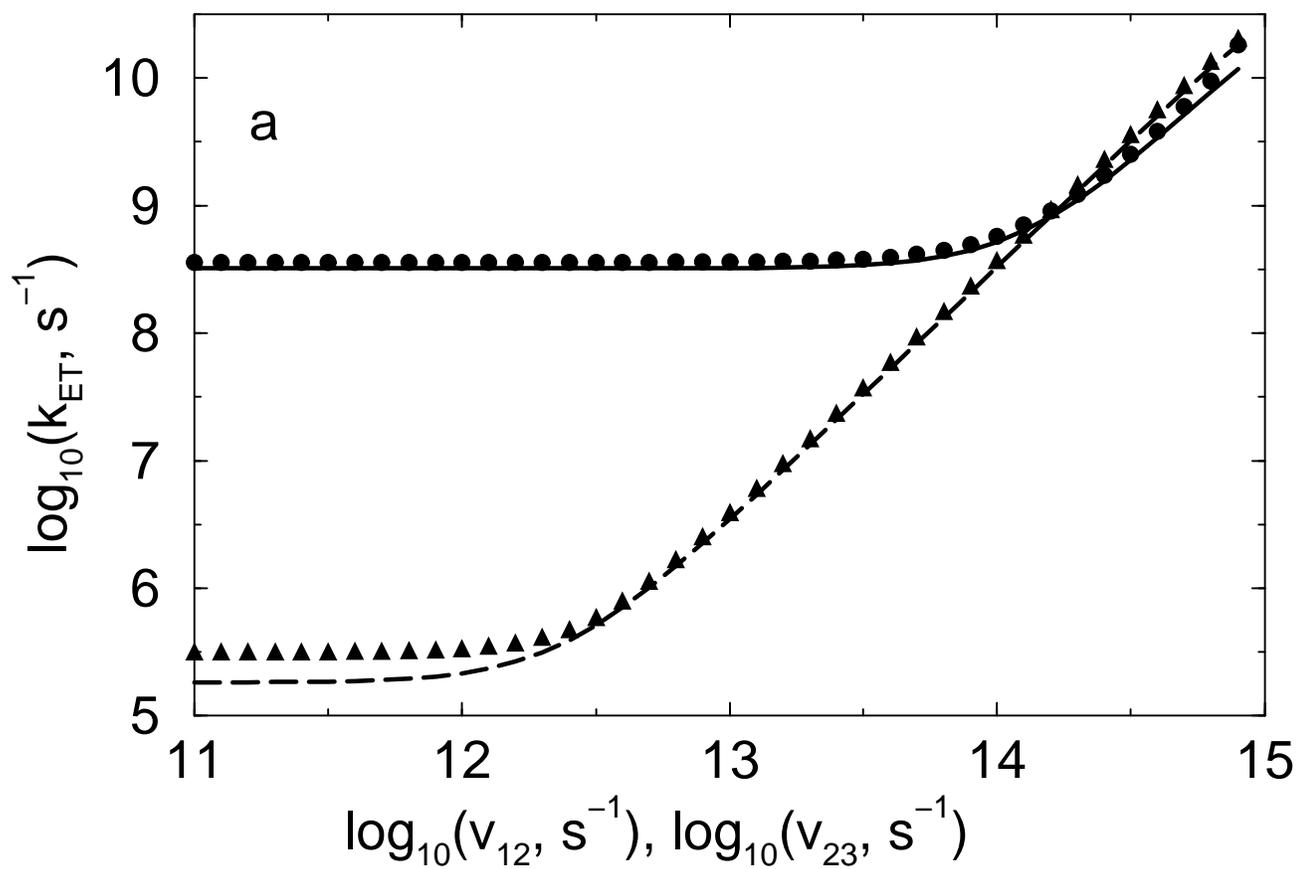
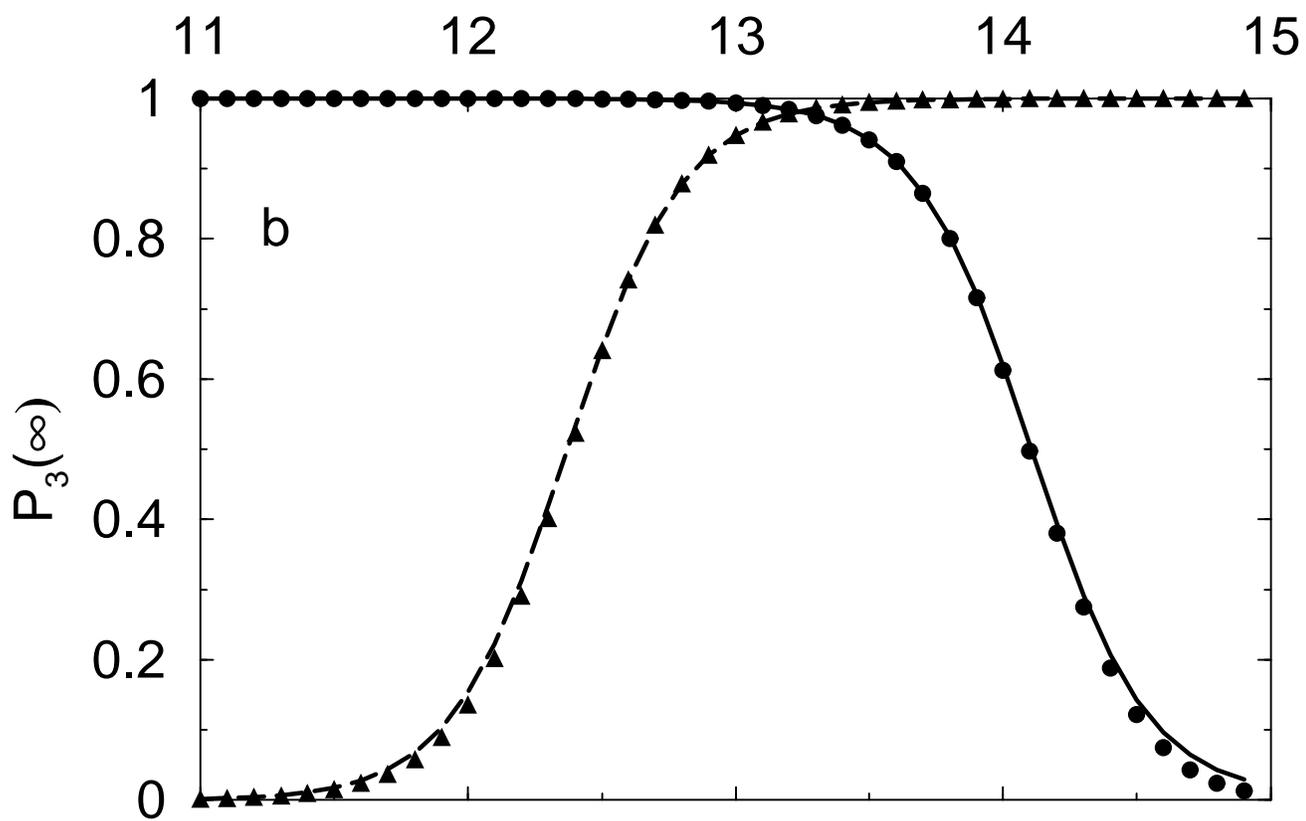
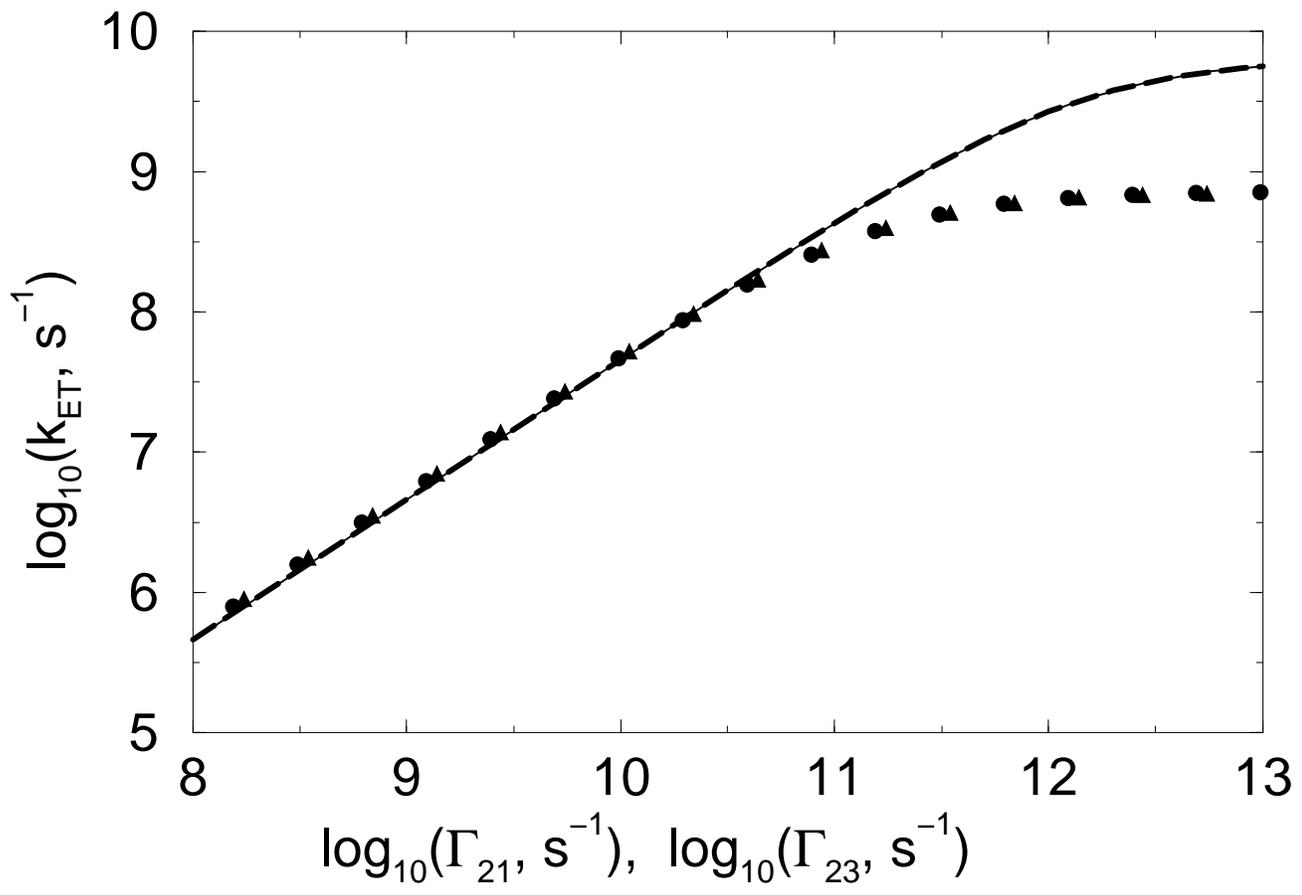
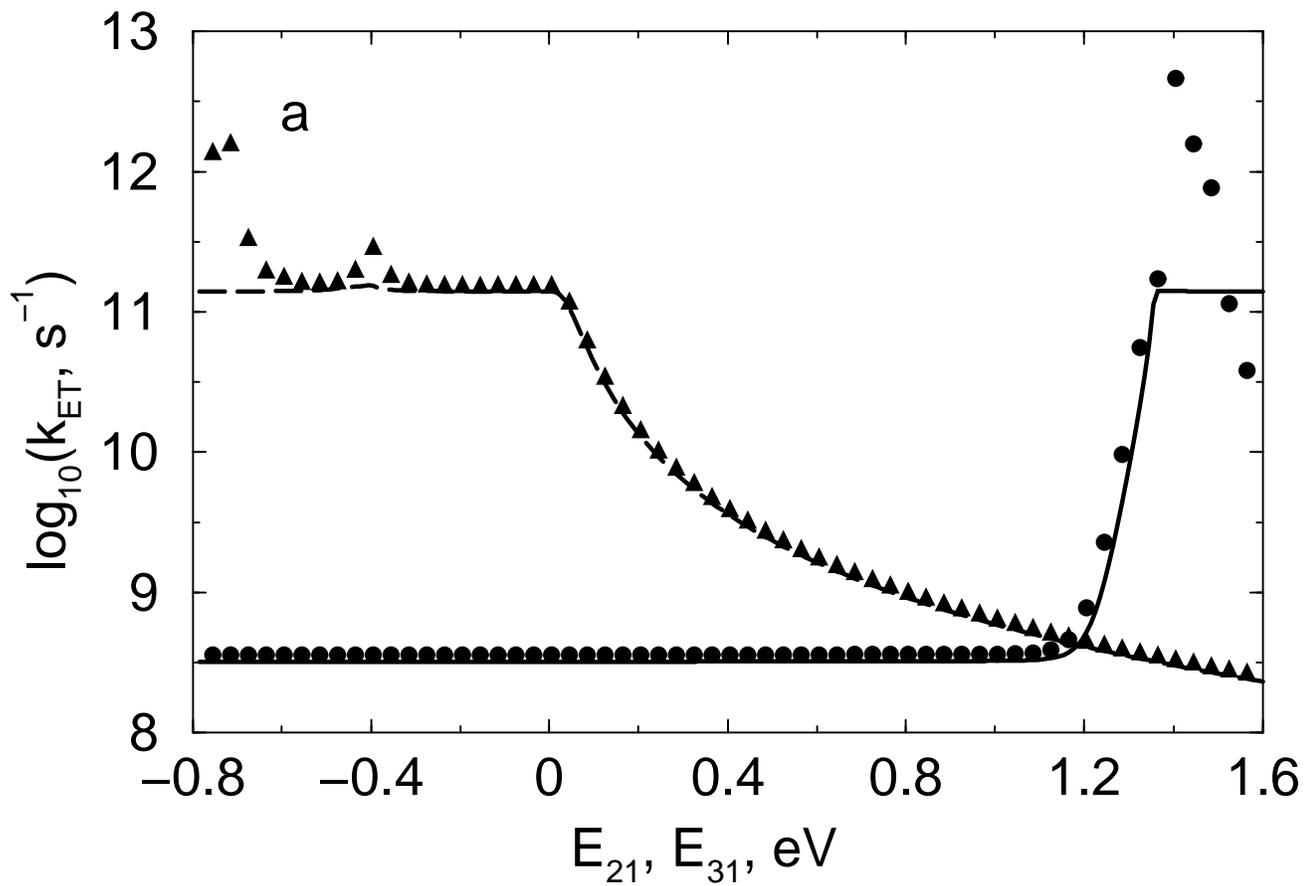
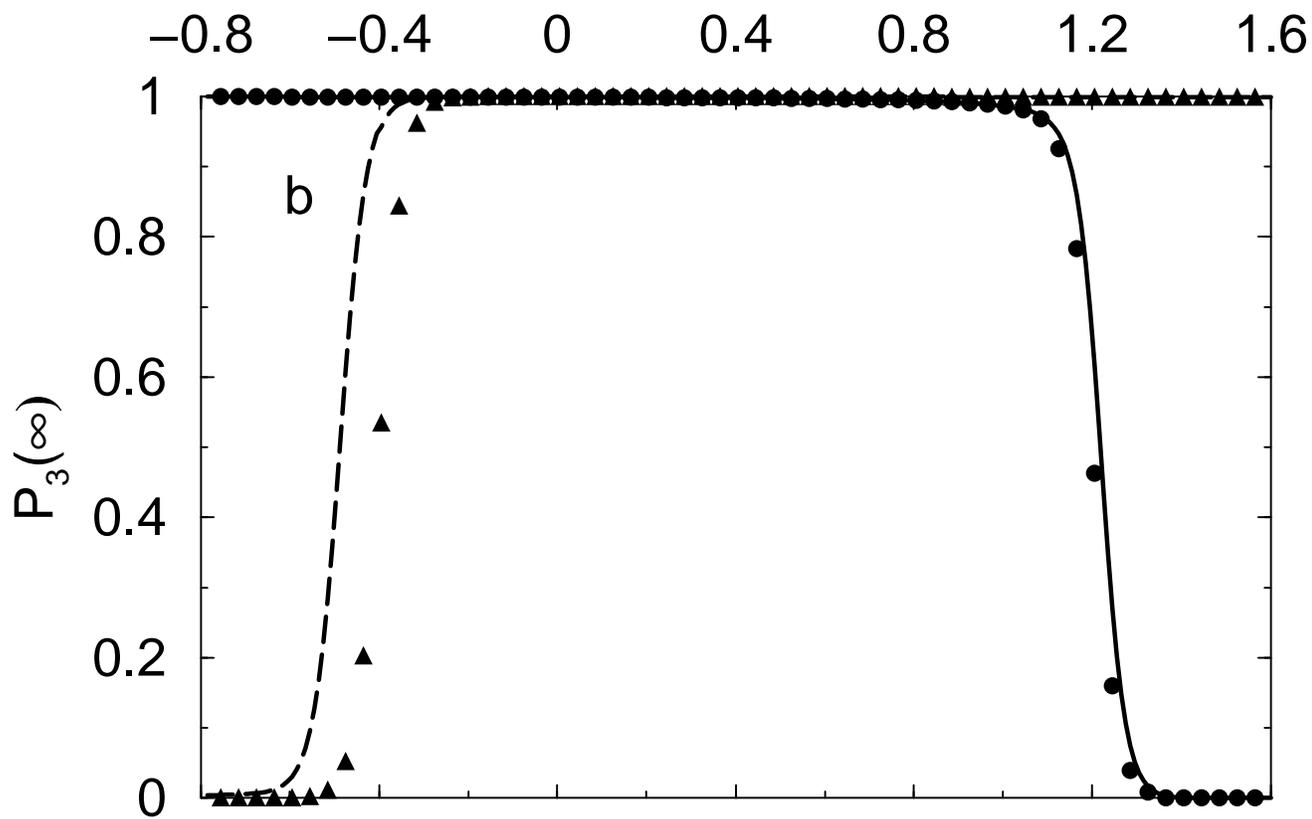
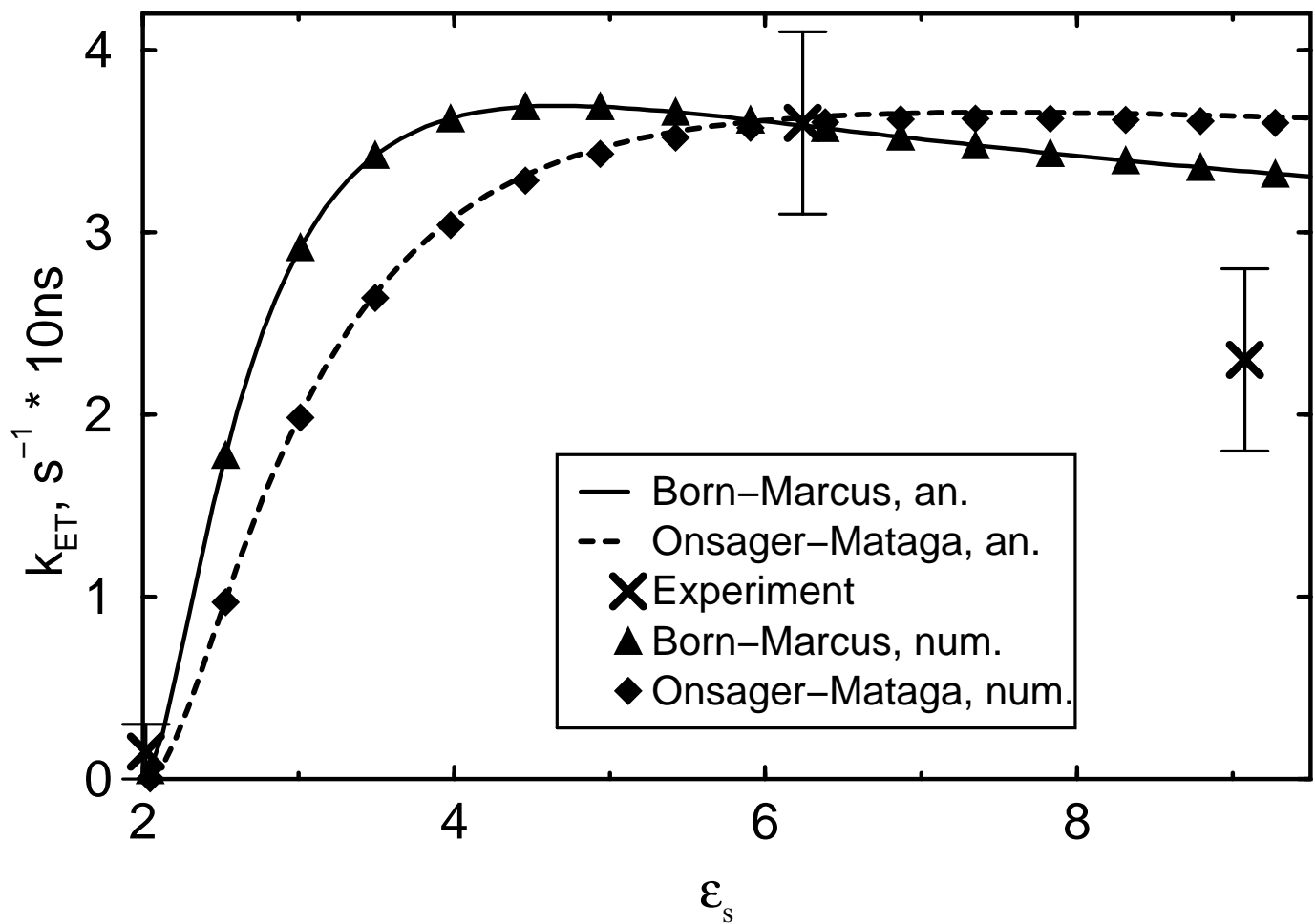
\end{document}